\documentclass[aps,prd,preprint,showkeys,showpacs]{revtex4-1}
\usepackage{graphicx}                                
\usepackage{amsmath,amsfonts,bbm}
\graphicspath{{./Figures/}}                          

\usepackage{epsfig}
\usepackage{amssymb}
\usepackage{amsfonts}
\usepackage{float}
\newcommand{\Tr}{\mbox{Tr}}

\newcommand{\beq}{\begin{equation}}
\newcommand{\eeq}{\end{equation}}
\newcommand{\bea}{\begin{eqnarray}}
\newcommand{\eea}{\end{eqnarray}}
\newcommand{\beas}{\begin{eqnarray*}}
\newcommand{\eeas}{\end{eqnarray*}}
\newcommand{\eq}{\begin{equation}}
\newcommand{\en}{\end{equation}}
\newcommand{\eqa}{\begin{eqnarray}}
\newcommand{\ena}{\end{eqnarray}}
\newcommand{\Fig}[1]{Fig.~\ref{#1}}

\newcommand{\Eq}[1]{Eq.~(\ref{#1})}

\begin{document}


\title{BRST symmetry vs. Horizon condition in Yang-Mills theories}
\author{G.~Burgio}
\author{M.~Quandt}
\author{H.~Reinhardt}
\affiliation{Institut f\"ur Theoretische Physik, Auf der Morgenstelle 14, 
72076 T\"ubingen, Germany}

\begin{abstract}
We show that in SU(2) Yang-Mills theories a simple relation exists between 
lattice gluon propagators in Coulomb and Landau gauge and discuss the physical 
implications of such result. In particular, the realization of the Gribov-Zwanziger 
confinement mechanism in Coulomb gauge, linked to dual-superconductivity, would 
imply that the standard BRST charge 
must be ill defined non perturbatively. As a consequence, the Kugo-Ojima 
confinement criterion, which relies on BRST charge conservation beyond
perturbation theory, would not be fulfilled.
\end{abstract}

\maketitle

\section{Introduction} 

\subsection{Overview}

Understanding confinement in non-Abelian gauge theories has proved a 
major challenge for theoretical physics. How to enforce Gau\ss's law, which 
ensures gauge invariance and
vanishing color charge on physical states, 
$Q_{c} |\Psi\rangle_{\mbox{\scriptsize ph}} = 0$; 
how to explain the discrepancy 
between color neutral S-matrix hadronic states and color charged fields 
appearing in the action; and how the IR 
properties of gauge fixed Green's functions should reflect confinement of gluons and 
quarks and non-physicality of ghost fields have been, from
the very beginning, closely 
intertwined questions propaedeutic to the appearance of a linearly rising 
potential and which have spawned extensive research dating back to the '70s.

In a seminal paper, Kugo and Ojima \cite{Kugo:1979gm} gave 
a criterion on the IR behavior of the ghost Green's function that ensures
S-matrix states being color neutral in 
{\it covariant} gauges with the Faddeev-Popov (FP) action and the
corresponding Becchi-Rouet-Stora-Tyutin (BRST) symmetry. According to their 
result, the ghost dressing function should diverge as the 4-momentum goes to 
zero. The key assumption underlying their derivation is that a global BRST 
charge can be defined beyond perturbation theory, such that 
$Q_{\mbox{\scriptsize BRST}} |\Psi\rangle_{\mbox{\scriptsize ph}} = 0$. 
Color neutrality of all S-matrix states and the disappearance of ghosts 
from the spectrum is then ensured by the so-called quartet mechanism 
\cite{Kugo:1979gm}. Equivalently, if 
$Q_{c} |\Psi\rangle_{\mbox{\scriptsize ph}} = 0$ and the 
Kugo-Ojima criterion is not fulfilled, BRST symmetry must be non-perturbatively
either spontaneously broken or an ill-defined concept.

A complementary approach was pioneered by Gribov \cite{Gribov:1977wm} and later
expanded by Zwanziger \cite{Zwanziger:1995cv}. The basic idea is that the 
FP mechanism is not sufficient to define the non-Abelian partition function
beyond perturbation theory. To extract a unique
field along the gauge orbit the functional integral needs to be restricted, 
via terms which are in general non-local and are often referred to as the 
``Horizon condition'', to the $1^{\mbox{st}}$ Gribov Region $\Omega$, where 
the FP operator is positive definite, or even further to the Fundamental Modular 
Region $\Lambda$, where the gauge fixing functional has only absolute minima. 
Obviously, any restriction on the integration domain implemented other than through
the standard FP mechanism will in general break BRST symmetry.
Although it can be applied also to covariant gauges, the 
physical implications of the Gribov-Zwanziger approach are transparent in 
{\it Coulomb} gauge. The latter has the advantage of projecting, in the Hamiltonian 
formulation, onto the subspace of states satisfying Gau\ss's law
\cite{Reinhardt:2008pr,Reinhardt:2008ij}, so that no additional assumptions
to ensure $Q_{c} |\Psi\rangle_{\mbox{\scriptsize ph}} = 0$,
as in the Kugo-Ojima approach, are needed
and an explicit construction of the gauge invariant Hilbert space 
as in Ref.~\cite{Burgio:1999tg} can be circumvented. 
The Horizon condition
implies in Coulomb gauge an IR diverging ghost dressing function and a 
vanishing
{\it static} transverse gluon propagator \cite{Gribov:1977wm}. Moreover since 
the 
inverse of the ghost dressing function can be identified with the 
dielectric function of the Yang-Mills vacuum \cite{Reinhardt:2008ek},
the latter behaves like a perfect
color dia-electric medium, i.e. a dual-superconductor; 
a confining potential can arise via dual-Meissner effect.

\subsection{Motivations}

Approximation schemes for the set of Dyson-Schwinger equations (DSE)
in QCD had been proposed already by Mandelstam in the late '70s 
\cite{Mandelstam:1979xd}, leading to the now abandoned hypothesis of infrared 
slavery. A first 
self consistent solution for the DSE was obtained by 
Cornwall \cite{Cornwall:1981zr} in the pinch technique framework
\cite{
Binosi:2009qm}, leading to a massive IR behavior of the gluon 
propagator. In such solution the Kugo-Ojima
criterion is not fulfilled and BRST symmetry is lost non perturbatively.
Only from the late '90s on alternative approaches
have been developed to find solutions to the tower of DSE satisfying BRST 
invariance \cite{vonSmekal:1997is}. Since then, advances in functional 
methods for the non perturbative analysis of Yang-Mills Green's functions in Landau 
\cite{Fischer:2006vf,Fischer:2008uz} and
Coulomb \cite{Szczepaniak:2001rg,Feuchter:2004mk} gauge on one side and
the availability of large lattice simulations with improved gauge fixing 
techniques 
\cite{Leinweber:1998uu,Bloch:2003sk,Bogolubsky:2005wf,Bogolubsky:2007bw,%
Cucchieri:2007rg,Burgio:2008jr,Bogolubsky:2009qb,Bornyakov:2009ug} on the 
other side have revived the interest in the IR analysis of gauge fixed Green's functions.

As mentioned above, in Landau gauge continuum methods allow in general for two 
types of solutions \cite{Fischer:2008uz}. When coupled to the 
requirement that a global BRST 
charge be defined and conserved, as assumed by the Kugo-Ojima confinement 
criterion for the Landau ghost dressing function, they
lead in any dimension to a conformal behavior of the gluon 
propagator in the IR, the 
so called ``scaling'' solution \cite{vonSmekal:1997is,Fischer:2006vf,Fischer:2008uz}. 
Imposing BRST symmetry to hold beyond the perturbative regime 
assumes implicitly that no extra care needs to be taken to restrict the functional 
integral beyond the FP operator. Relaxing this requirement, however, 
the gluon propagator can acquire a massive IR behavior 
\cite{Cornwall:1981zr,Dudal:2008sp,Fischer:2008uz,Kondo:2009ug}
while the ghost dressing function stays finite, also known
as ``decoupling'' solution in the literature.
In principle BRST symmetry gets explicitly broken when rendering local the terms 
in the action enforcing the Horizon condition, since this introduces additional ghost 
fields and induces modified residual transformations. Such terms should
lead to a massive IR gluon propagator \cite{Dudal:2008sp}. Alternatively, it has 
been argued in the literature
that a vanishing gluon propagators can still be obtained in this framework
\cite{Zwanziger:1992qr,Zwanziger:2009je,Zwanziger:2010iz} and that 
dimensional two condensates 
\cite{Akhoury:1997by,Burgio:1997hc,Boucaud:2000ey} are eventually responsible
for the massive IR solutions as in Ref.~\cite{Dudal:2008sp}.

As for the lattice, all results for the gluon and ghost in Landau gauge obtained 
in ``standard'' 
setups clearly favor a IR massive gluon in 2+1 and 3+1 dimensional Yang-Mills 
theory, while agreement with scaling is only found in 1+1 dimensions 
\cite{Cucchieri:2007rg,Cucchieri:2008fc}. 
Some works applying ``non-standard''  
lattice methods have however pointed to possible discrepancies in such picture
\cite{Silva:2005hb,Sternbeck:2008mv,Maas:2009se,Maas:2009ph,Oliveira:2009nn}.

Whether one or the other scenario
is realized in QCD is the subject of an ongoing controversy. It should be noticed 
that in both cases the gluon propagator
violates reflection positivity and disappears from the physical spectrum anyhow
\cite{Cucchieri:2004mf}. Moreover, a linearly rising potential could arise in
both frameworks \cite{Braun:2007bx}. Indeed, the issue boils down to whether color 
charge descends from BRST charge or, equivalently, whether in Landau gauge the 
Horizon condition can be neglected in favor of  the standard FP mechanism. 

In  Coulomb gauge we are in a more comfortable position: in 3+1 dimensions 
both lattice results \cite{Burgio:2008jr} and functional methods 
\cite{Feuchter:2004mk,Epple:2006hv} show a 
IR vanishing static gluon transverse propagator
well described by Gribov's formula \cite{Gribov:1977wm}, while the
Coulomb ghost dressing function clearly diverges at zero momentum 
\cite{Voigt:2007wd,Quandt:2008zj}. The Gribov-Zwanziger mechanism is
realized and a coherent 
picture emerges, naturally enforcing $Q_{c} |\Psi\rangle_{\mbox{\scriptsize ph}} = 0$ 
and with a direct physical interpretation in terms of dual Meissner effect 
\cite{Reinhardt:2008ek}. As we will show below, 
the 2+1 dimensional Yang-Mills theory behaves analogously. 

As for the 1+1 dimensional case, the theory is topological 
\cite{Witten:1988hf}
and no physical gluon exists. As a ``back of the envelope'' calculation 
shows, the Coulomb gauge propagator 
vanishes identically in momentum space and the corresponding FP operator can 
be inverted exactly, while the Landau propagator only 
describes pure gauge degrees of freedom \cite{Birmingham:1991ty}.
An IR conformal behavior with 
conserved BRST charge is thus natural \cite{Witten:1988hf,Birmingham:1991ty}. 
Indeed, topological theories are the starting points for the construction of 
non-perturbative BRST invariance in any
dimensions. Interestingly enough, lattice results, which 
implement some sort of Horizon condition 
restricting by construction to $\Omega$, and continuum calculations using only
the FP mechanism seem to coincide in this case.

In higher dimensions however it is not clear whether the Horizon condition 
in Landau gauge can be reconciled with a vanishing gluon propagator. 
As mentioned above, different arguments have been brought up in favor 
\cite{Zwanziger:1992qr,Zwanziger:2009je,Zwanziger:2010iz} and against 
\cite{Cornwall:1981zr,Dudal:2008sp,Kondo:2009ug} it.
Conventional lattice results, including direct simulations of the Kugo-Ojima
function \cite{Furui:2004cx}, are quite unambiguous in favoring the latter approach,
although gauge fixing ambiguities and/or problems in the definition
of the lattice gauge fields have been invoked to claim possible
agreement with the former scenario 
\cite{Sternbeck:2008mv,vonSmekal:2008es,Maas:2009ph}, though 
the massive gluon IR behavior turns out to be very robust even when 
choosing ad-hoc
prescriptions for the gauge fixing procedure to enhance the ghost propagator 
\cite{Maas:2009se}. It is also
not clear at the moment how a redefinition of the gauge fields 
\cite{vonSmekal:2008es} would 
circumvent the well known no-go theorem for a lattice BRST symmetry
\cite{Neuberger:1986vv,Neuberger:1986xz}. 
Even if such approach would work, it would anyhow imply a deep change of
paradigm for lattice gauge fixing, since on one hand a BRST invariant construction 
would sum over all copies in the different Gribov regions while
on the other hand the standard gauge fixing procedure on the lattice follows
Gribov's spirit, effectively 
restricting the configurations to lie within $\Omega$. There is of course a residual
ambiguity in the choice of the best among all local minima found, since
the explicit restriction to $\Lambda$ would 
require an infinite numerical precision, but on one side
going to the continuum limit makes the gauge ambiguity much milder 
\cite{Bogolubsky:2009qb}, while 
on the other side thermodynamic arguments suggest that the relevant
contribution to the functional integral will anyway lie on 
$\partial \Lambda \cap \partial \Omega$, so that gauge ambiguities, apart
from topological ones, should not play any r\^ole. 

Since i) reliable physical input through phenomenological data is still lacking; ii) the 
matter turns out difficult 
to solve by brute force numerics, i.e. by explicitly going to the thermodynamic 
and/or continuum limit and iii) still waiting for a valid and viable BRST
construction on the lattice, we follow here an independent approach. 
In dimensions higher than 1+1 a physical, although confined, gluon obviously 
exists and one should be able to read its degrees of freedom out 
of both the transverse Landau and Coulomb propagators: a relation between the 
two and eventually to the physical 
gluon degrees of freedom should exist. Moreover if Gribov's prescription to restrict
the functional integral is the physically sound one in Coulomb gauge, since it seems
to agree with dual superconductivity, one would expect the Gribov mass 
appearing there \cite{Gribov:1977wm,Burgio:2008jr}
to be linked to some physical quantity, i.e it should effectively be gauge invariant. 
This mass should show up in Landau, interpolating \cite{Cucchieri:2007uj}
and covariant gauges \cite{Cucchieri:2009kk} and be related to the
effective IR gluon mass \cite{Cornwall:1981zr}.

In $d$+1 dimensions the most natural Ansatz to relate the static transverse
gluon propagator in Coulomb gauge and the transverse gluon propagator
in Landau gauge is via a one to one correspondence between 
the variables
they must be function of, namely $\vec{p}^2$, the natural O($d$) invariant in the static 
Hamiltonian Coulomb gauge picture, and $p^2$, its O($d$+1) covariant counterpart in
euclidian lattice Landau gauge. We therefore assume that in 2+1 and 3+1 
dimensions by just relating monotonically the 
$d$-and $d+1$-momentum scales the two propagators will coincide. 
Based on this simple Ansatz we will show that i) there is a one to one 
correspondence between the lattice results for the IR vanishing static Coulomb 
gluon propagator, 
realizing Gribov's original Horizon condition \cite{Gribov:1977wm} and the 
IR massive Landau propagator and ii) that within this correspondence the 
Coulomb gauge Gribov mass exactly coincides with the Landau gauge
effective IR gluon mass. Should the uniqueness
of the massive behavior for lattice gluon propagator in Landau gauge
be confirmed, one could interprete the loss of BRST symmetry 
beyond perturbation theory that would descend from it as a consequence of the 
dual-superconductor mechanism in QCD.

\section{Results} 
We calculate the transverse gluon propagator in Coulomb and Landau gauge in 
2+1 and 3+1 dimensional SU(2) Yang-Mills theory on configurations generated via
MC simulations on lattices up to $64^3$ and $32^4$ and $\beta$ between 3-12 
and 2.15-2.6 respectively, using a heath bath plus over-relaxation algorithm. 
The scale is set re-expressing the lattice spacing in units of the string 
tension \cite{Bloch:2003sk,Teper:1998te}, which we fix to 
$\sigma=(440\, {\rm MeV})^2$ also in 2+1 dimensions.

\subsection{Coulomb propagator} 
\label{couprop}
In Coulomb gauge renormalization issues on the lattice
have only been recently clarified in Ref.~\cite{Burgio:2008jr,Burgio:2008yg}.
We extend here our procedure to 2+1 dimensions and improve our gauge fixing 
algorithm. 

For each fixed $\beta$ we define the gluon propagator $D^{ab}_{ij}(p)$ 
as the Fourier transform of the gluon two-point function:
\beq
D^{ab}_{ij}(p) =
\left\langle \widetilde{A}^a_{i}({k}) \widetilde{A}^b_{j}(-{k})
\right\rangle_U
   = \delta^{ab} \delta_{ij} D_\beta(p)\;, \qquad  
p_{\mu} = p({k}_{\mu}) = \frac{2}{a} \sin\left(\frac{\pi
      {k}_{\mu}}{L}\right)\;.
\label{eq:D-def}
\eeq
Here $\widetilde{A}^a_{\mu}({k})$ is the Fourier transform of the lattice 
gauge
potential $A^a_\mu(x+\hat{\mu}/2)$, defined as:
\beq
A_{\mu}(x+ {\hat \mu}/2) =
       \frac{1}{2i}\left(U_{\mu}(x)-U_{\mu}(x)^{\dagger}\right)\;,
\label{eq:gp}
\eeq
$a$ is the lattice spacing, $p=(\vec{p},p_0)$ denotes the four-momentum and
$~{k}_{\mu} \in (-L/2, +L/2]$ are the integer-valued lattice momenta.
We always select 
spatial indices
$\vec{k}$ satisfying a cylinder cut \cite{Leinweber:1998uu,Voigt:2007wd} to 
minimize 
violations of rotational invariance. Time-like momenta $p_0$ are unconstrained.

For each MC generated configuration we first fix one of the $2^d$ 
flip sectors \cite{Bogolubsky:2007bw}, where $d$ are the spatial dimensions, 
and then fix Coulomb gauge 
maximizing separately for every time slice $t$ the gauge functional:
\beq 
F_g(t) = \frac{1}{6 L^3} \sum_{\vec{x},i} \mbox{Tr} ~
U^g_{i}(\vec{x},t)\;,\qquad
U^g_{i}(\vec{x},t) = g(\vec{x},t) ~U_{i}(\vec{x},t) 
~g^{\dagger}(\vec{x}+{\hat{i}},t)
\label{eq:functional}
\eeq
with respect to local gauge transformations $g(\vec{x},t) \in$ SU(2),
employing a simulated 
annealing plus over-relaxation algorithm.
The 
local maxima of $F_g(t)$ satisfy for each fixed $t$ the differential 
lattice Coulomb gauge transversality condition for the gauge potentials:
\beq
\partial_{i} A^g_{i}(\vec{x},t) =
       A^g_{i}(\vec{x}+{\hat i}/2,t)-A^g_{i}(\vec{x}-{\hat{i}}/2,t) = 0\;.
\label{eq:transversality}
\eeq
We choose the best $F_g(t)$ out of $n_c$
random gauge copy for each time slice $t$ separately, where usually $n_c=5$, 
and combine 
them into the best copy of the whole configuration for the flip sector chosen.
We now go to the next sector and repeat the procedure. At the end we
chose the sector with the highest global functional $F_g = \sum_t F_g(t)$
as the best copy of the configuration.
The number of total recursions is $2^d\, L\, n_c$. Increasing $n_c$ leads
of course to a better gauge fixing. However, going to higher 
volumes, one might consider to increase $n_c$ only for the sectors with
the highest $F_g$ \cite{Bogolubsky:2007bw}.

To extract the $p_0$ dependence we need to fix the time
gauge $g(t)$. 
We choose here the integrated Polyakov gauge prescription proposed in
Ref.~\cite{Burgio:2008jr,Burgio:2008yg}: 
\beq
u(t)\,=\frac{1}{L^3}
\sum_{\vec{x}} U_0(\vec{x},t) \to {\rm const.}
\label{eq:6}
\eeq 
corresponding in the continuum to
\beq
0=\int\, d^3x\, \partial_\mu A_\mu(\vec{x},t) \Rightarrow \partial_0 \int\, 
d^3x\, A_0(\vec{x},t) = 0
\label{eq:6a}
\eeq 
To achieve this, we define 
\beq
\hat{u}(t) = \frac{u(t)}{\sqrt{{\mbox{Det}}\left[u(t)\right]}} \in SU(2)\,.
\eeq
Periodic boundary conditions in $t$ make
$\Tr \prod_t \hat{u}(t) = \Tr\,P$ invariant under $g(t)$.
\Eq{eq:6} can thus be fixed recursively through
\beq
g(t) \hat{u}(t) g^{\dagger}(t+1) = \tilde{u}\equiv P^{{1}/{L}}
\eeq
up to a global gauge $g(0)$ satisfying $\left[g(0),P\right]=0$. We can choose 
$g(0)=\mathbbm{1}$. 

The main observation, made in Ref.~\cite{Burgio:2008jr,Burgio:2008yg} for the 
3+1 dimensional case and
which we find here to hold in 2+1 dimensions as well, is that the 
bare gluon propagator 
factorizes as:
\beq
D_\beta(|\vec{p}|,p_0) = \frac{f_\beta(|\vec{p}|)}{|\vec{p}|^2}
\frac{{g}_\beta(z)}{1+z^2}\,,\qquad z=\frac{p_0}{|\vec{p}|}\;.
\label{eq:fact}
\eeq
The function $g_\beta(z)$ will in general depend on the temporal
gauge $g(t)$. However the 
static propagator, defined as
\beq
D_{\beta}(|\vec{p}|) = \int_{-\infty}^\infty \frac{d p_0}{2 \pi} 
D_\beta(|\vec{p}|,p_0)=  
\frac{{f}_\beta(|\vec{p}|)}{\pi |\vec{p}|}\int_0^\infty d z
\frac{g_\beta(z)}{1+z^2}\;,
\label{eq:int}
\eeq 
will be independent of such choice. 
As shown in Ref.~\cite{Burgio:2008jr,Burgio:2008yg} lattice cutoffs to the 
integral in \Eq{eq:int} cause
scaling violations; as a viable \cite{Nakagawa:2009is}
alternative to the
lattice Hamiltonian limit \cite{Burgio:2003in}
we define $D_{\beta}(|\vec{p}|)$ directly from $f_\beta(|\vec{p}|)$, 
extracting it from 
the data after dividing out $g_\beta(z)$ and averaging over $p_0$:
\beq
D_{\beta}(|\vec{p}|):= \frac{{f}_\beta(|\vec{p}|)}{|\vec{p}|}\propto 
|\vec{p}|\, \sum_{p_0} D_\beta(|\vec{p}|,p_0)\frac{1+z^2}{{g}_\beta(z)}\;.
\label{eq:enind}
\eeq 
We obtain in this way a multiplicative renormalizable $D_{\beta}(|\vec{p}|)$, 
$D_{\mu}(|\vec{p}|)= Z(\beta,\mu) D_{\beta}(|\vec{p}|)$. We choose 
$\mu=\infty$,
i.e. we fix the overall normalization such that 
$\lim_{|\vec{p}|\to\infty} |\vec{p}| D_{\mu}(|\vec{p}|) =1$.

\begin{figure}[htb]
\mbox{
\includegraphics
{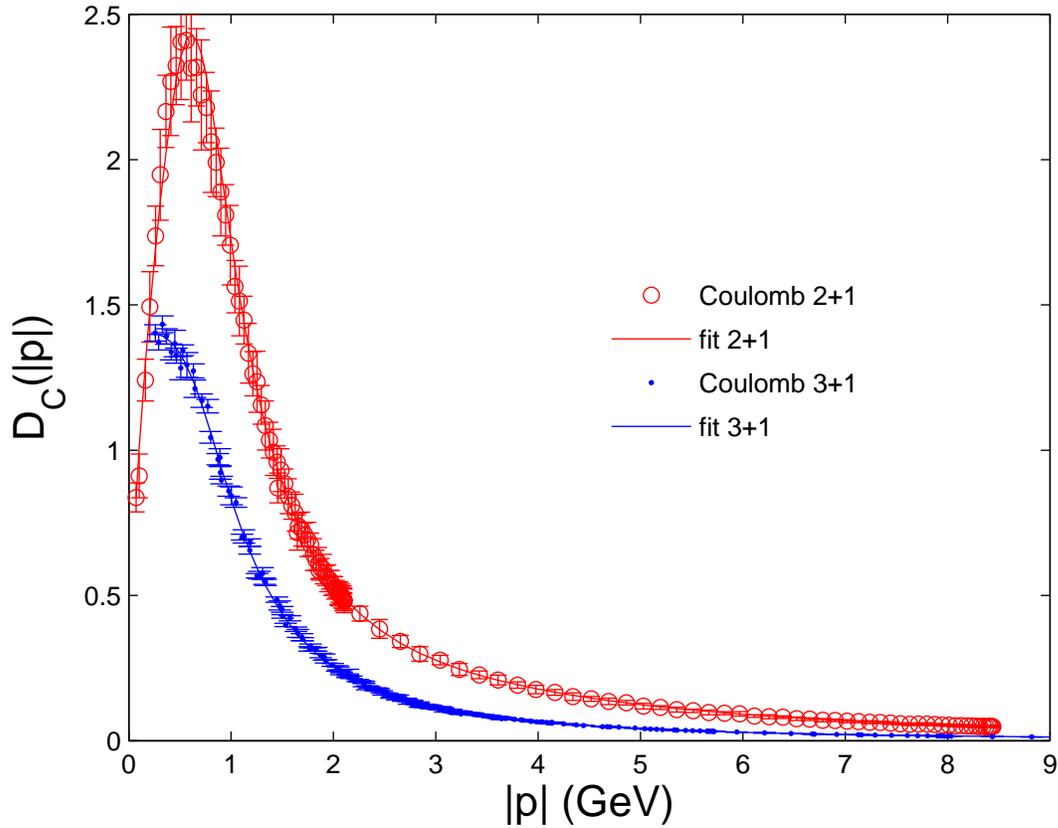}
}
\caption{${D}_{C}(|\vec{p}|)$ and fits in 2+1 and 3+1 dimensions.}
\label{fig1}
\end{figure}

In 2+1 dimensions a Gribov like IR leading behavior:
\beq
G(|\vec{p}|) = \frac{|\vec{p}|}{\sqrt{|\vec{p}|^{4} + M^4}}\,,
\label{eq:gribov}
\eeq
plus corrections of the type:
\beq
\tilde{G}(|\vec{p}|)= 
\frac{|\vec{p}|^{1+\alpha}}{\sqrt[\beta]{|\vec{p}|^{(2+\alpha)\beta} + \mu^{(2+\alpha)\beta}}}
\label{eq:c2+1}
\eeq 
describes the data extremely well. To account for the intermediate momentum 
region different
choices for the sub-leading terms other than \Eq{eq:c2+1} can be made, 
the only constraints being
asymptotic behaviors $|\vec{p}|^{-1}$ in the UV and $o(|\vec{p}|)$ in the 
IR. Lacking any theoretical input further restricting the functional form
is beyond our scope.
In \Fig{fig1} we show  
$D_{C}(|\vec{p}|)=|\vec{p}|^{-1}D_{\mu}(|\vec{p}|)$, which
nicely extrapolates to a constant in the IR,
together with the fit to $|\vec{p}|^{-1}(c\,G(|\vec{p}|)+
\tilde{c}\, \tilde{G}(|\vec{p}|))$. To improve readability versus the
3+1 dimensional propagator we have 
normalized the data to $3.5 |\vec{p}|^{-2}$ in the UV.
We find $M = 4.6(4)$ GeV and 
 $0.2\leq \chi^2$/d.o.f. $\leq 0.6$, depending on the choice of the 
sub-leading terms, our main source of error.

In 3+1 dimensions $D_{\mu}(|\vec{p}|)$ is well described by Gribov's 
formula alone \cite{Burgio:2008jr}.
In \Fig{fig1} we 
show $D_{C}(|\vec{p}|)=|\vec{p}|^{-1}D_{\mu}(|\vec{p}|)$ 
and its fit to $|\vec{p}|^{-1}G(|\vec{p}|)$ as in \Eq{eq:gribov}; 
we obtain $M=0.856(8)$ GeV
with $\chi^2$/d.o.f. = 1.6. In Ref.~\cite{Burgio:2008jr} the deep IR data 
slightly 
deviated from a constant behavior, resulting in a $\chi^2$/d.o.f. = 3.3
for the same fit. These 
deviations are here, with increased volume and improved
gauge fixing, considerably smaller, halving the $\chi^2$/d.o.f. value.
We therefore decided to ignore possible sub-leading corrections as 
in \Eq{eq:c2+1} for the time being. Only a study on larger volumes and closer 
to the continuum limit, in the spirit of the Landau 
gauge analysis of
Ref.~\cite{Cucchieri:2007md,Bogolubsky:2009qb,Bornyakov:2009ug}, could clarify
the situation. We also stress that no logarithmic corrections are
necessary to describe the data in the UV. Although this is expected in 2+1 
dimensions,
it is a highly non trivial result in the 3+1 dimensional case, since
one loop calculations for $D_\beta(|\vec{p}|,p_0)$ give an explicit anomalous 
dimension 
\cite{Watson:2007mz}. However, contrary to covariant 
gauges, neither higher orders are under control nor a leading-log 
re-summation is viable. Cancellations when going from the one loop
full propagator to the full static propagator are therefore in principle possible.

As a last remark, both in 2+1 and 3+1 dimensions the asymptotic behaviors of 
$D_\mu(|\vec{p}|)$ 
are in agreement with the results
of the IR and UV  analysis of the DSE in the Hamiltonian approach 
of Ref.~\cite{Feuchter:2004mk,Feuchter:2007mq}.
\subsection{Landau propagator} 
The lattice gluon propagator in Landau gauge $D_L(p)$ has been the subject of 
extensive research; we refer the interested
reader to the relevant literature regarding restoration of
rotational invariance, gauge copy effects, continuum and thermodynamic limits
etc. \cite{Leinweber:1998uu,Bloch:2003sk,Bogolubsky:2005wf,Bogolubsky:2007bw,%
Cucchieri:2007rg}. Here we simply apply the by now well
established methods developed in the above references. In particular, 
we basically follow Ref.~\cite{Bogolubsky:2005wf,Bogolubsky:2007bw}. 

\begin{figure}[htb]
\mbox{
\includegraphics
{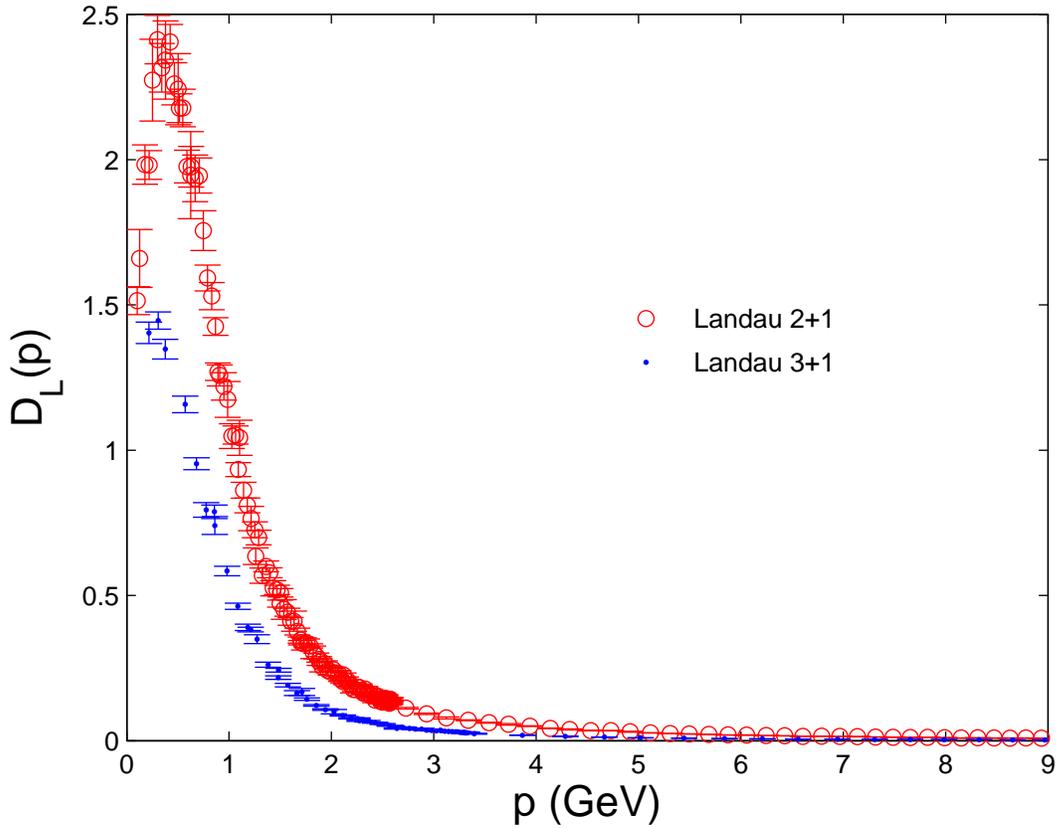}
}
\caption{$D_L(p)$ in 2+1 and 3+1 dimensions.}
\label{fig2}
\end{figure}

The results for 2+1 and 3+1 dimensions, normalized
such that the peaks coincide with those of \Fig{fig1}, are shown in 
\Fig{fig2} and agree with all known previously published results.

\subsection{Landau vs. Coulomb} 

\begin{figure}[htb]
\mbox{
\includegraphics
{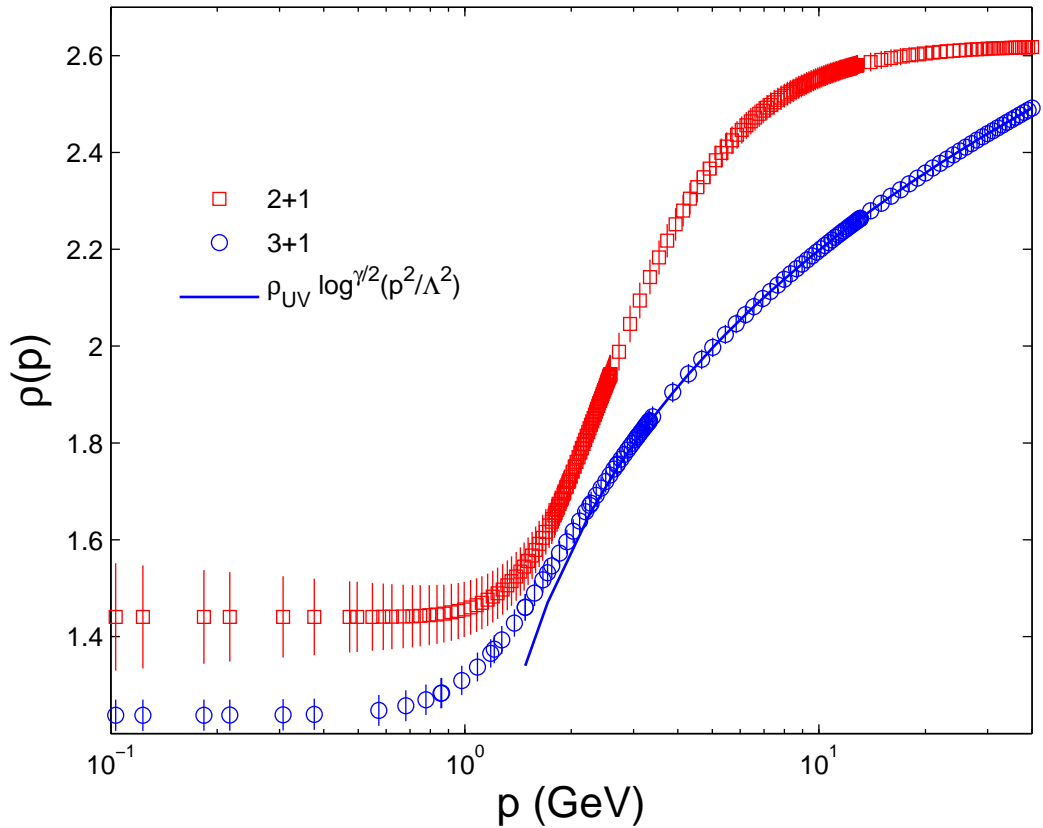}
}
\caption{$\rho(p)$ in 2+1 and 3+1 dimensions, the latter compared to the 
perturbative asymptotic behavior.
}
\label{fig3}
\end{figure}

\begin{figure}[htb]
\mbox{
\includegraphics
{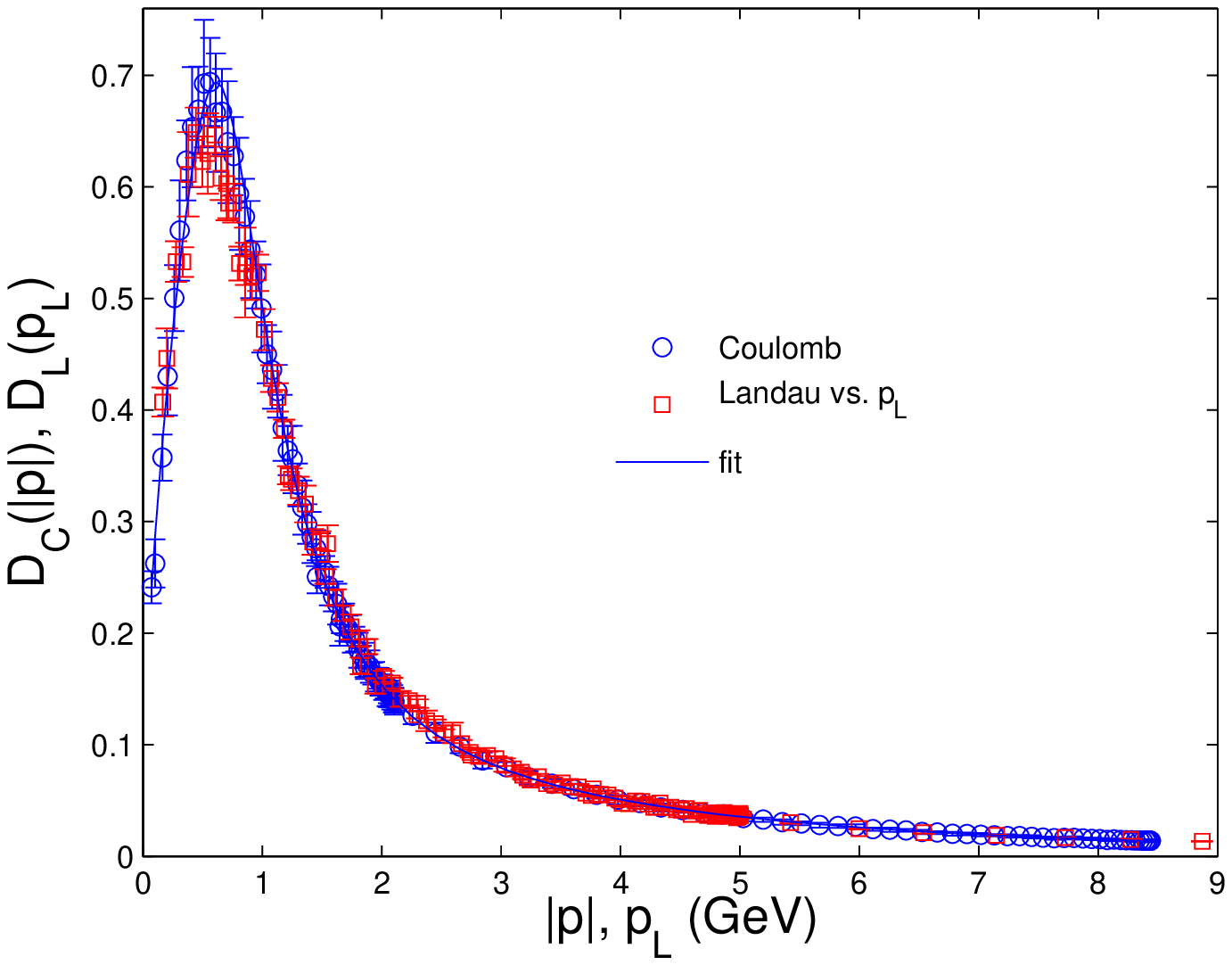}
}
\caption{${D}_{C}(|\vec{p}|)$, $D_L(p_L(p))$ and fit in 2+1 dimensions.}
\label{fig4}
\end{figure}

\begin{figure}[htb]
\mbox{
\includegraphics
{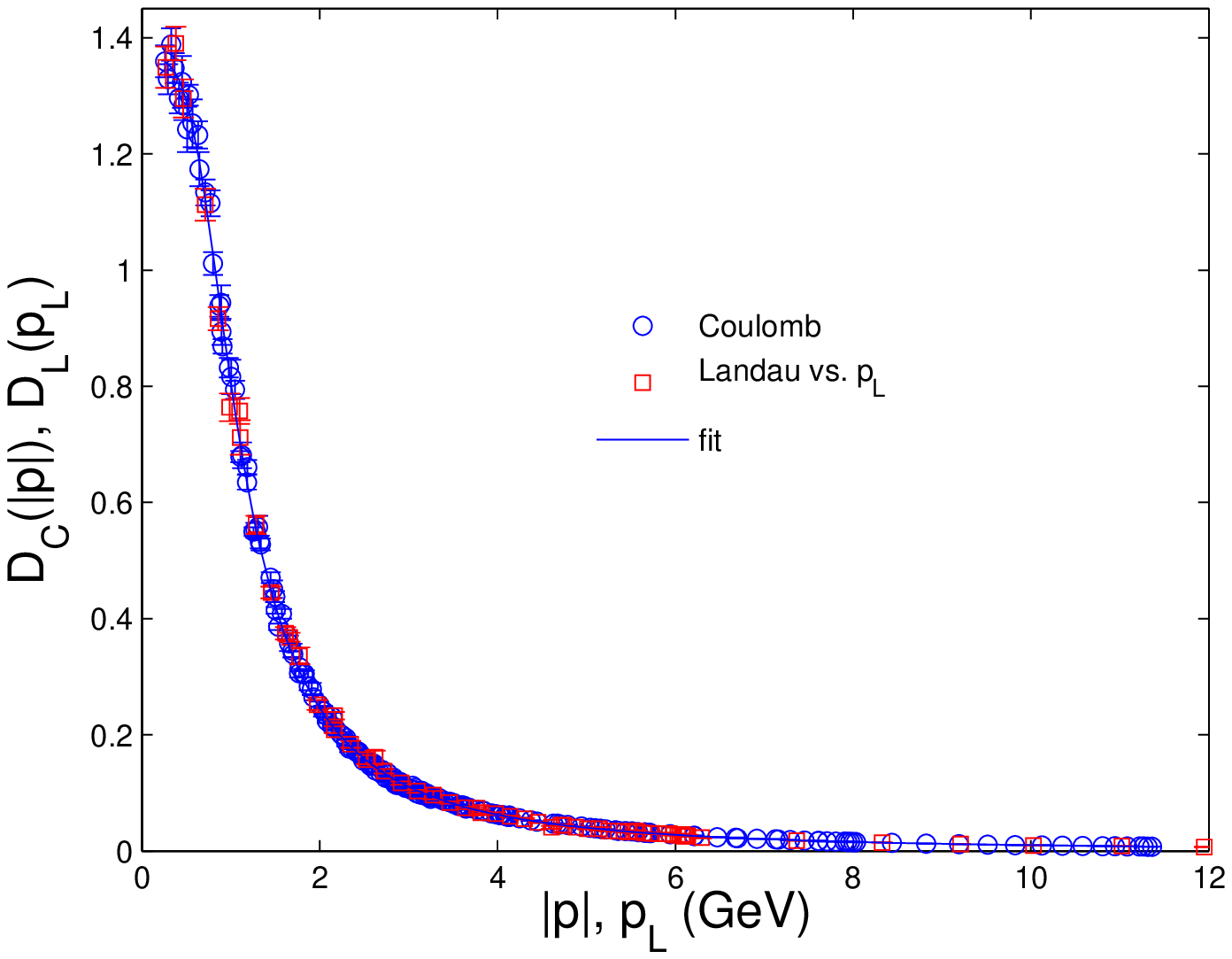}
}
\caption{${D}_{C}(|\vec{p}|)$, $D_L(p_L(p))$ and fit in 3+1 dimensions.}
\label{fig5}
\end{figure}

On trivial dimensional grounds, based on their tree level UV behavior,
it is obvious that $D_L(p)$ should be compared to 
${D}_{C}(|\vec{p}|)=|\vec{p}|^{-1}D_{\mu}(|\vec{p}|)$.
For both 2+1 and 3+1 dimensions a simple inspection of the Landau gauge 
transverse propagator plot against $p=\sqrt{|\vec{p}|^2+p_0^2}$ 
given in \Fig{fig2} shows striking similarities
with ${D}_{C}(|\vec{p}|)$
plotted against ${|\vec{p}|}$, as in \Fig{fig1}.
The two curves almost coincide, up to a faster fall off of $D_L(p)$ 
in the UV. In fact, it simply looks as if the Landau 4-momentum scale $p$
is stretched by some factor when compared to the Coulomb momentum scale 
$|\vec{p}|$. For a quantitative comparison, motivated by this observation,
we make the simple Ansatz $D_L(p) \equiv {D}_{C}(p_L(p))$, where
$p_L(p) =  p\, \rho(p/\Lambda)$ is strictly monotonous and $\rho(p/\Lambda)$ is a 
dimensionless function
interpolating between two constants $\rho_{\mbox{\scriptsize IR}}$ 
and $\rho_{\mbox{\scriptsize UV}}$ in 2+1 dimensions, where the
anomalous dimension vanishes, and between a constant 
$\rho_{\mbox{\scriptsize IR}}$ and 
$\rho_{\mbox{\scriptsize UV}} 
\log^{\gamma/2}{p^2/\Lambda_{\mbox{\scriptsize QCD}}^2}$ in 3+1 dimensions. 
We use for $\rho(p/\Lambda)$ a ratio of rational functions of the same order,
$\rho(p/\Lambda) = R_n(p^2/\Lambda^2)/R_d(p^2/\Lambda^2)$, corrected in 
3+1 dimensions by $\log^{\gamma/2}{R_l(p^2/\Lambda_{\mbox{\scriptsize QCD}}^2)}$, 
$R_l$ being also a rational function. Polynomials of degree $\leq 3$
for $R_{n,d,l}(x)$ are sufficient for our scope. In 3+1 dimensions 
shuffling the $p$ dependence inside or outside the $\log$ is somewhat 
arbitrary. Indeed, a form of the type $\rho(p/\Lambda_{\mbox{\scriptsize QCD}}) =  
\rho_{\mbox{\scriptsize IR}} \log^{\gamma/2}{R_l(p^2/\Lambda_{\mbox{\scriptsize QCD}}^2)}$ 
is already sufficient, so that in both 2+1 and 3+1 dimensions one single
mass parameter, $\Lambda$ or $\Lambda_{\mbox{\scriptsize QCD}}$ is needed to 
describe $\rho$. A certain level of 
arbitrariness is also present in the normalization of the highest and lowest coefficient
of $R_{n,d,l}(x)$ as well as in their degree. The errors quoted below reflect all such, 
partly correlated, uncertainties, which will in particular show up in the value of 
$\Lambda$ and $\Lambda_{\mbox{\scriptsize QCD}}$.
However, lacking further theoretical input,
pinning down $\rho(p)$ to a precise functional form goes beyond the scope 
of this work. We have fitted both $|\vec{p}|{D}_{C}(|\vec{p}|)$ and 
$p_L(p) D_L(p_L(p))$ at the same time to 
\Eq{eq:gribov} (plus \Eq{eq:c2+1} in 2+1 dimensions). $\rho(p)$ as obtained
from the fits is shown in \Fig{fig3}, while the resulting comparison between
${D}_{C}(|\vec{p}|)$ vs. $D_L(p)$ plotted against $p_L(p)$ in 2+1 and 3+1 
dimensions is given in \Fig{fig4} and \Fig{fig5}.
Numerically, we obtain 
$\rho_{\mbox{\scriptsize IR}} = 1.44(14)$,
$\rho_{\mbox{\scriptsize UV}}=2.610(15)$, $\Lambda=3.0(3)$ GeV and 
$\chi^2$/d.o.f. = 1.2 in 2+1 dimensions, 
$\rho_{\mbox{\scriptsize IR}} = 1.255(15)$, 
$\rho_{\mbox{\scriptsize UV}}=1.49(1)$, 
$\gamma = 0.52(3)$ and $\Lambda_{\mbox{\scriptsize QCD}} = 1.05(15)$ GeV 
with $\chi^2$/d.o.f. = 1.6 in 
3+1 dimensions; the fitted values for all parameters in 
\Eq{eq:gribov} and \Eq{eq:c2+1}, particularly for $M$, exactly
coincide with those of Section~\ref{couprop}. Moreover our 3+1 dimensional value 
for $M=0.856(8)$ GeV is in excellent 
agreement with recent SU(3) estimates of the IR mass in Landau gauge which 
explicitely include a dimensional two condensate in the analysis \cite{Dudal:2010tf}.

It is obvious that there
is no way to relate ${D}_{C}(|\vec{p}|)$ to a scaling, IR vanishing $D_L(p)$ while also 
accounting for the UV region, since a monotonous mapping between the momenta 
could never do the job. One could of course object that the choice, in our opinion 
natural, to compare the Landau 
propagator with ${D}_{C}(|\vec{p}|)$ already shuts the doors to any scaling behavior.
One could therefore attempt to compare 
$D_L(p)$ directly to $D_{\mu}(|\vec{p}|)$ again via a monotonous mapping of the 
momenta, e.g. $D_L(p) \equiv D_{\mu}(\bar{p}_L(p))$, to check if an agreement
to the same level of accuracy as in our analysis could be reached.
In this case the function $\bar{p}_L(p)$ should behave {\it up to
dimensionful constants} like $\propto p^\kappa$ in the IR
and $\propto p^2$ (plus logarithms in 3+1 dimensions) in the UV, 
i.e. $\bar{p}_L(p)$ would in principle need two new mass scales, besides
the one for the intermediate momentum region, to account for the IR and UV behavior. 
As discussed above and as can be inferred from \Fig{fig3}, this should be 
compared to the present analysis, where the only mass scale needed in 
$\rho(p)$ basically coincides with the Gribov mass both
in 2+1 and 3+1 dimensions, even though its exact value will depend on the,
still arbitrary, parameterization of $\rho(p)$. Moreover, that the Gribov 
parameter $M$ extracted both from the lattice Coulomb and 
Landau gauge gluon transverse degrees of freedom coincide to such 
degree of precision constitutes in our opinion a highly non trivial result supporting a 
physical interpretation for the dynamically generated gluon mass. 
Of course the points raised above do not forbid in principle the existence of a 
mapping between the Gribov-like Coulomb propagator and a scaling like solution
in Landau gauge, but lacking any theoretical arguments and as long as one can
neither obtain a scaling solution on the lattice nor a solution obtained with 
continuum methods is available to sufficient degree of precision, any analogous 
quantitative check as the one performed in this paper to prove or disprove such 
correspondence will remain impossible.


\section{Conclusions}
We have shown that in 2+1 and 3+1 dimensional Yang-Mills theory, 
where physical gluons propagate, a natural and simple relation exists between the 
lattice results for the gluon-gluon Green's functions in two different gauges, i.e. the 
IR finite
transverse Landau propagator $D_L(p)$ and the IR vanishing static transverse 
Coulomb propagator ${D}_{\mu}(|\vec{p}|)$. Indeed, the two are equivalent if one
simply rescales the momenta. i.e. $D_L(p) = 
p_L(p)^{-1} {D}_{\mu}(p_L(p))$ where
$p_L(p) = p\, \rho(p)$ and $\rho(p)$ is given in \Fig{fig3}. 
${D}_{\mu}(|\vec{p}|)$ agrees with the Gribov-Zwanziger scenario and has a 
natural interpretation 
within the dual-superconductor picture of confinement \cite{Reinhardt:2008ek}.
This corroborates a monopole condensation interpretation of the
gluon massive solution in Landau gauge as well, making, as long as no scaling 
like solution can be found on the lattice, the non
perturbative loss of BRST symmetry more bearable.
Whether the static Coulomb gluon itself coincides with the Yang-Mills field's 
physical degrees of freedom is still an open issue 
\cite{Chen:2008ag,Chen:2009mr}. 
Should the uniqueness of the IR massive gluon behavior on the lattice be confirmed,
it would suggests that in covariant gauges in more than one spatial dimension
the Gribov-Zwanziger \cite{Gribov:1977wm,Zwanziger:1995cv} term should be 
explicitly included in the action, BRST symmetry would be non-perturbatively broken 
and the Kugo-Ojima criterion \cite{Kugo:1979gm} could not be fulfilled.
Indeed an effectively massive gluon propagator is natural in
a theory where monopoles condense, as can be explicitly seen in the 
Georgi-Glashow model \cite{Polyakov:1976fu,Dunne:2000vp};
dual-superconductivity is expected to work in a similar 
way to confine gluons, albeit with a dynamically generated 
scalar boson \cite{'tHooft:1974qc}. Indeed, a topological origin for dynamic
mass generation in QCD has been argued long time ago in 
Ref.~\cite{Cornwall:1979hz} and offers a natural interpretation
of the pinch technique result \cite{Cornwall:1981zr,Binosi:2009qm}.
In topological theories on the other hand, where the physical Coulomb 
propagator 
should identically vanishes, a IR conformal confinement with a non perturbatively
conserved BRST charge can be realized. The 
Chern-Simons and BF theories will be interesting cross-checks to confirm 
the different mechanism between topological and QCD like theories. 

\section*{Acknowledgments} 
We would like to thank Peter Watson and Jan M. Pawlowski for 
stimulating discussions. 
This work was partly supported by DFG under 
the contract DFG-Re856/6-3. 
\bibliographystyle{apsrev}
\bibliography{references}

\end{document}